\documentclass[%
reprint,
amsmath,amssymb,
aps, groupedaddress,superscriptaddress, nofootinbib
]{revtex4-2}
\usepackage[usenames,dvipsnames]{xcolor}
\usepackage{amssymb}
\usepackage{graphicx}
\usepackage{amsmath}
\usepackage{comment}
\usepackage{bbm}
\usepackage[bookmarks=true,colorlinks,citecolor=blue,urlcolor=blue]{hyperref}
\usepackage{braket}
\usepackage{babel}
\usepackage{blindtext}
\usepackage[normalem]{ulem}
\usepackage{physics}

\usepackage{mathtools}

\definecolor{mypurp}{rgb}{0.35, 0, 0.7}

\usepackage{amsthm}
\usepackage{txfonts}

\theoremstyle{definition}

\def\hL{\mathcal{L}}

\begin{document}
\def\papertitle{{Dissipative phase transitions and passive error correction}}

\title{\papertitle}

\newcommand{\TUM}{\affiliation{Technical University of Munich, TUM School of Natural Sciences, Physics Department, 85748 Garching, Germany}}
\newcommand{\MCQST}{\affiliation{Munich Center for Quantum Science and Technology (MCQST), Schellingstr. 4, 80799 M{\"u}nchen, Germany}}

\def\umd{Joint Quantum Institute, NIST/University of Maryland, College Park, Maryland 20742, USA}
\def\qcs{Joint Center for Quantum Information and Computer Science, NIST/University of Maryland, College Park, Maryland 20742, USA}
\def\aws{AWS Center for Quantum Computing,  Pasadena, California 91125, USA}

\author{Yu-Jie Liu} 
\TUM \MCQST

\author{Simon Lieu}
\thanks{This work was done prior to joining AWS.}
\affiliation{\umd}
\affiliation{\qcs}
\affiliation{\aws}

\date{\today}
\begin{abstract}

We  classify different ways to \textit{passively} protect classical and quantum information, i.e.~we do not allow for syndrome measurements, in the context of local Lindblad models for spin systems.  Within this family of models, we suggest that passive error correction is associated with 
nontrivial phases of matter  and propose a definition for dissipative phases based on robust \textit{steady state} degeneracy of a Lindbladian in the thermodynamic limit. We study three thermalizing models in this context: 
the 2D Ising model, the 2D toric code, and the 4D toric code. In the low-temperature phase, the 2D Ising model hosts a robust classical steady state degeneracy while the 4D toric code hosts a robust \textit{quantum} steady state degeneracy.  We perturb the models with terms that violate detailed balance and observe that qualitative features  remain    unchanged, suggesting that $\mathbb{Z}_2$ symmetry breaking in a Lindbladian is useful to protect a classical bit while intrinsic topological order protects a qubit.

\end{abstract}

\maketitle

\section{Introduction} 

One of the central challenges toward building a practical quantum computer is the ability to correct quantum errors \cite{nielsen_book, lidar_book, gottes1997}.  Most error correcting schemes  that are currently being pursued rely on  redundantly encoding logical information into many physical qubits, and constantly measuring stabilizer operators (via two-qubit gates and ancilla qubits) to ensure that logical information remains uncorrupted \cite{bravyi:1998,eric2002, fowler:2012}. Fast and accurate measurements pose a significant experimental challenge and come with severe hardware overhead.  A central question therefore remains to identify different ways to correct quantum errors that do not rely on syndrome measurements.

A notable alternative goes by the name \textit{passive}  quantum error correction \cite{lidar_book}: The thermal bath associated with certain Hamiltonians naturally leads to dissipative processes that  correct thermal errors \cite{eric2002, bacon2006, Alicki2009, Alicki2010, loss2010,yoshida2011, haah2011, pastawski2011, bravyi2013,  bombin2013,  terhal2015,  bombin2015, breuckmann2016b, williamson2016, brown2016,    kubica2023b, li2023c}.  A prominent example is the 4D toric code \cite{eric2002,Alicki2010}, which has the property that
quantum information  initially encoded in the ground state can be recovered at any finite time, if the temperature of the bath is below a critical value (in the thermodynamic limit).

In this work, we would like to understand such passive  correction through  the lens of phase transitions in  local, Markovian systems. Formulating the problem in this way allows us to: (1) classify generic mechanisms that produce a qubit steady state structure in the thermodynamic limit, (2) suggest that detailed balance (i.e.~thermal equilibrium) is not crucial for the passive error correcting properties of the bath; rather, the locality and symmetry of perturbations  are important in determining the stability, (3) draw parallels with the field of \textit{driven-dissipative} phase transitions, which  can also  exhibit error correcting properties but do not rely on thermal equilibrium.

We classify different mechanisms that result in a robust steady state degeneracy of the Lindbladian (i.e. multiple steady states). While finite-sized systems can host such degeneracies \cite{albert2014, prosen2012}, they are typically fragile to arbitrary local perturbations. We therefore focus our attention on models that have an exponentially-good degeneracy only in the thermodynamic limit. We suggest  that $\mathbb{Z}_2$ spontaneous symmetry breaking in a Lindbladian leads to a robust classical bit. Prominent examples of this include  the thermal Ising model in 2D \cite{eric2002} and the driven-dissipative cat code \cite{mirrahimi2014,  amazon_cat, lieu2020}. We also suggest that  intrinsic topological order in a Lindbladian (i.e.~robust steady state degeneracy) appears in the low-temperature phase of the 4D toric code.

The connection between passive error correction and dissipative phase transitions is then apparent: The former  \textit{requires}  a robust steady state degeneracy, while the latter is \textit{characterized} by  such a degeneracy \cite{lieu2020}. This is reminiscent of  phase transitions in closed quantum systems, where the thermodynamic limit of a nontrivial phase is typically characterized by a stable ground state degeneracy \cite{wen_book}.

Beyond potential applications to error correction, our work also sheds light on the nature of topological phase transitions in dissipative systems. The field of dissipative phase transitions has focused primarily on studying spontaneous symmetry breaking \cite{diehl2008,  maghrebi2016,young2020, keeling2013, rossini2018,  ciuti2018, lieu2020, eisert2017, cirac2012, minganti2023}. In this work, we suggest that Lindbladians can also undergo   \textit{topological} phase transitions, characterized by a robust steady state degeneracy and a  closing of the dissipative gap at the phase boundary. 
Our analysis points to   fundamental questions  regarding topological phase transitions in open systems,  which we discuss in the outlook.

\section{Quantum memory in the thermodynamic limit} \label{sec:qm} Consider a Hilbert space $\mathcal{H}$, and define two encoded, logical states $| \bar 0 \rangle,| \bar 1 \rangle \in\mathcal{H}$ that span the codespace $\mathcal{C}$. We assume the system is always initilized in the codespace: $\rho_i = | \psi \rangle \langle \psi |$ where $\ket{\psi}\in\mathcal{C}$. In this paper, the code space will be the degenerate ground state subspace of a Hamiltonian.

A continuous-time Markovian generator $\mathcal{L}$ in Lindblad form is defined by
\begin{align} \label{eq:lindblad}
    \frac{d \rho}{dt}  = \hL(\rho) = -i[H,\rho] +\sum_j  \left(L_j\rho L_j^{\dag}-\frac{1}{2}\{L^{\dag}_j L_j,\rho\}\right),
\end{align}
where $H$ is the Hamiltonian of the system and $L_j$ are  dissipators that arise due to the system-environment coupling \cite{lindblad1976}. For the thermal baths considered in this work, we can split the Lindbladian into two contributions that occur with different rates, a zero-temperature part and an infinite temperature part:
\begin{equation}
    \mathcal{L}_T = \kappa_0 \mathcal{L}_{T=0}  +\kappa_\infty \mathcal{L}_{T=\infty} .
\end{equation}
The temperature is determined by the ratio of these two processes ($\kappa_0, \kappa_\infty$) and both processes are local in space. The zero-temperature contribution $\mathcal{L}_0$ represents the corrections that  send the system to the code space (ground state manifold). The  $\mathcal{L}_\infty$ contribution represents errors that cause the state to leave the code space (i.e.~bit flips and phase flips). We suppose that this noisy process occurs for a finite time $t$ that sends  $\rho_i$ to a mixed state $\rho_m(t) = e^{\mathcal{L} t}  (\rho_i)$.

One of the prerequisites for a quantum memory is that the superoperator $\mathcal{L}$ needs to have  degenerate (i.e.~more than one) steady states in the presence of noise. (If $\mathcal{L}$ has a unique steady state then  arbitrary qubit initial states will become indistinguishable on a time scale of order the inverse dissipative gap.) In particular, $\mathcal{L}$ needs to have at least four eigenvalues of zero: Two to protect the relative populations between the logical states, and two to protect the relative complex phase. If the state preserves quantum information, it can be expressed in the following form: 
\begin{equation} \label{eq:noiseless}
\rho_m(t)= e^{\mathcal{L} t}  (\rho_i) = \left(
\begin{matrix}
|c_0|^2 & c_0 c_1 \\
 c_0^* c_1^* & |c_1|^2
\end{matrix}
\right) \otimes M(t),
\end{equation}
where $M$ is a diagonal matrix that does not have to be pure: $\text{Tr}[M^2] \leq 1$. For a mixed $M$, such a structure is called a ``noiseless subsystem'' \cite{lidar_book}. We will show explicit examples of such steady states that have diagonal $M$ matrices that follow the Boltzmann distribution. In a fully passive error correcting scheme, one imagines directly manipulating the quantum information that is stored in the mixed state above,  then only doing one destructive measurement of the qubits at the end of the computation, avoiding the need to measure stabilizers throughout the protocol.

It is theoretically convenient to  quantify the decay rate of coherences by allowing for  a ``single-shot'' decoding superoperator $\mathcal{E}$ which sends every state in the Hilbert space back to the code space according to some algorithm (e.g.~minimal weight matching for the toric code) \cite{kyungjoo2021b}.   The final state we end up with is:
\begin{equation}
\rho_f(t) = \mathcal{E} e^{\mathcal{L} t} (\rho_i).
\end{equation}
We wish to find generic setups where the difference between the initial and final state  is   exponentially small in the system size for any arbitrary (but finite) $t$. Specifically, we will focus on the deviation of the  overlap from unity:
\begin{equation}
1- \text{Tr}[\rho_i \rho_f(t)]  \sim e^{-c N},
\end{equation}
where $N$ is the system size and $c$ is a constant.  The continuous-time Markov process  is capable of ensuring that the errors do not  destroy the  quantum information.

What are the generic conditions under which a continuous-time Markovian generator $\mathcal{L}$ will host a noiseless subsystem   steady state structure of the form of Eq.~\eqref{eq:noiseless}? 
In this work, we  investigate   emergent noiseless subsystems that only appear in the \textit{thermodynamic limit} of  a $\mathbb{Z}_2$-symmetry-broken phase, or a phase with intrinsic topological order.  This differs from most examples in the literature, where exact noiseless subsystems arise in finite systems due to non-Abelian symmetries \cite{Zhang_2020}.  For such systems, the qubit will decohere in the presence of either local bit-flip errors ($X$) or local phase-flip errors ($Z$) (see Appendix~\ref{sec:finite}).  By contrast $\mathbb{Z}_2$ symmetry breaking will protect against local  bit flips but not  phase flips, and intrinsic topological order will protect against both. This is summarized in Table \ref{tab:noiseless}.  This closely mirrors quantum phase transitions: A nontrivial quantum phase supports a \textit{robust} ground state degeneracy of the Hamiltonian in the thermodynamic limit. For $\mathbb{Z}_2$ spontaneous symmetry breaking, the degeneracy is fragile to terms which violate the   $\mathbb{Z}_2$ symmetry,  while for intrinsic topological order any local perturbation cannot split the degeneracy. 

\begin{figure}
    \centering
    \includegraphics[scale=0.45]{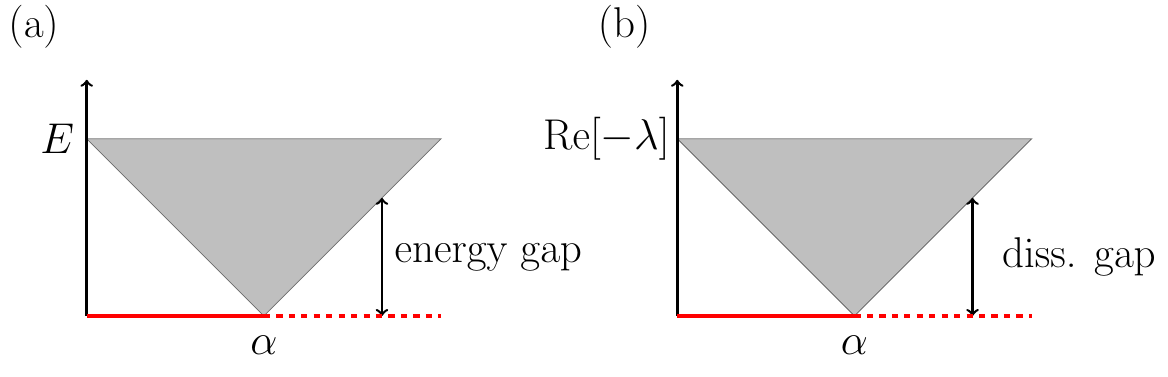}
    \caption{  (a) Caricature  spectrum of a Hamiltonian  $H(\alpha)$ across a   quantum phase boundary. In a nontrivial phase the ground state is degenerate (red line); the energy gap closes at a critical point, then a unique ground state emerges in the trivial phase (dashed red line). (b) We are looking for analogous phase transitions in a Lindbladian $\mathcal{L}(\alpha)$, characterized by a steady-state degeneracy in a nontrivial phase, and a closing of the dissipative gap at the phase boundary. }
    \label{fig:gap_fig}
\end{figure}

Working by analogy, we will search for   phase transitions in a Lindbladian  $\hL(\alpha)$ as we continuously deform some parameter $\alpha$ (e.g.~the temperature). In particular, we require that:
\begin{itemize}
\item{ $\mathcal{L}$ is composed of local terms.}
\item{ $\mathcal{L}$ has a  steady state degeneracy (e.g.~noiseless subsystem) in the thermodynamic limit of a nontrivial phase. This degeneracy can  be removed when  going across a phase boundary via smoothly tuning parameters $\alpha$ in $\mathcal{L}(\alpha)$.}
\item{ The steady state degeneracy is robust against arbitrary local perturbations in the master equation. (Up to symmetry constraints for spontaneous symmetry-breaking.) }
\item{ $\mathcal{L}$  has a non-zero dissipative gap away from the critical point and is gapless at the critical point \footnote{In a nontrivial phase, there are eigenvalues of the Lindbladian that are exponentially close to zero in system size, necessary for steady state degeneracy. We refer to  the gap as the smallest real part of the Lindblad spectrum above these steady state solutions, which should be finite at large system sizes for ``gapped'' systems.}. }
\end{itemize}
This last condition ensures that there is a critical slowing down of fluctuations at the   critical point:  
\begin{equation}
\langle  \mathcal{O}(t) \mathcal{O}   \rangle - \langle  \mathcal{O}   \rangle^2 \sim t^{- z},
\end{equation}
where $\mathcal{O}$ is an arbitrary observable, expectation values are taken with respect to the steady state, and $z$ is a dynamical critical exponent, i.e.~temporal correlators decay as a power-law rather than an exponential \cite{kardar2007}.   
Fig.~\ref{fig:gap_fig} provides a sketch of the requirements outlined above.

\begin{table}
\begin{center}
\begin{tabular}{c|c|c}
  noiseless subsystem & thermo. limit? & stable to noise? \tabularnewline
\hline
\hline 
non-Abelian strong symmetry  & no & no  \tabularnewline
\hline 
$\mathbb{Z}_2$ strong symmetry breaking &  yes & $X$   \tabularnewline
\hline
intrinsic topological order &  yes & $X$ and $Z$
\end{tabular}
\end{center}
\caption{ Different ways to achieve a noiseless subsystem  steady state structure. [See Eq.~\eqref{eq:noiseless}.] Imposing a non-Abelian strong symmetry \cite{Zhang_2020} on a \textit{finite} system ensures that the Lindbladian has a noiseless subsystem steady state but it is generally fragile to both bit flips ($X$) and phase flips ($Z$). (See Appendix~\ref{sec:finite}.) $\mathbb{Z}_2$ strong symmetry breaking \cite{lieu2020} requires the thermodynamic limit but is able to protect against bit flips (or phase flips, depending on convention). Intrinsic topological order requires the thermodynamic limit but is stable to both bit and phase flips. } 
\label{tab:noiseless}
\end{table}

In searching for a Lindbladian with the properties outlined above, it is useful to notice that the both symmetry-broken phases and topological phases can be thermally stable, and that the  Lindbladian can be used to describe  \textit{thermal} phase transitions  in these systems.
   The low-temperature phase of the former is useful for a  passive classical bit, while the latter is useful for a qubit \cite{eric2002}. In this work, we  study local dissipative models that  reproduce thermal phase transitions  in the 2D  Ising model, the 2D toric code,  and the 4D toric code. We perturb the models with terms that break detailed balance and observe that important features of the phase remain preserved. We provide  evidence that our model for the 2D   Ising model is an example of a symmetry-breaking phase transition that satisfies all of the bullet points above while the 4D toric code  is an  example of a topological transition with those properties. 

Beyond drawing  conceptual parallels between phase transitions in open and closed quantum systems, our  work raises the possibility that the dissipative 4D toric code is protected against \textit{arbitrary} local perturbations in the master equation.

\section{Two-dimensional Ising model}


We begin by considering spins on an $N \times N$ lattice. The   2D Ising model Hamiltonian reads
 \begin{equation}
 H_{is} = - \sum_{x,y = 1}^N (S_{x,y; r} + S_{x,y; t}),
 \end{equation}
 where
\begin{equation}
    S_{x,y; r} = Z_{x,y} Z_{x+1,y}, \qquad  S_{x,y; t} = Z_{x,y} Z_{x,y+1},
\end{equation}
are  $S_{x,y; r,t}$ are stabilizers  which pair up a spin on site $(x,y)$ with its neighbor to the right/top, $r,t$; $Z_{x,y}$ is the $Z$ Pauli operator on that site. The ferromagnetic states are the ground states of this model and span the code space: $| \bar{0} \rangle = | \uparrow \uparrow \uparrow \cdots \rangle, | \bar{1} \rangle = | \downarrow \downarrow \downarrow  \cdots\rangle$.

Let us define ``zero-temperature'' jump operators for each spin $x,y$: 
\begin{align} \label{eq:ising_diss }
L_{x,y}^{(4)} &= \sqrt{\kappa} X_{x,y} P_{x,y;r}^- P_{x,y;t}^- P_{x-1,y;r}^- P_{x,y-1;t}^- \\
L_{x,y}^{(3)} &=  \sqrt{ \tilde{\kappa}}  X_{x,y} P_{x,y;r}^+ P_{x,y;t}^- P_{x-1,y;r}^- P_{x,y-1;t}^- \\ 
L_{x,y}^{(2)} &=  \sqrt{  \kappa}  X_{x,y} P_{x,y;r}^+ P_{x,y;t}^+ P_{x-1,y;r}^- P_{x,y-1;t}^-, 
\label{eq:ising_diss2 }
\end{align}
where $\kappa,\tilde{\kappa}$  are the dissipative rates and  $P_{x,y;r/t}^{\pm} = (1 \pm S_{x,y;r/t})/2 $ is a projector onto a particular stabilizer configuration \cite{lieu2023b}. 
The superscripts indicate the number of domain walls  that  the projector is checking for (and we neglect to write jumps related by rotational invariance, e.g.~there are four different $L_{x,y}^{(3)} $ operators). These jump operators  
will only cause  a spin flip if it is  energetically favorable to do so.
We will also  consider uniform bit flips and phase flips on  each lattice site:
\begin{equation} \label{eq:ising_dep}
L_{x,y}' = \sqrt{\Delta_x} X_{x,y}, \qquad L_{x,y}'' = \sqrt{\Delta_z} Z_{x,y}.
\end{equation}
The dissipators in Eq.~\eqref{eq:ising_dep} represent the infinite temperature bath $\mathcal{L}_\infty$, while the ones defined in Eqs.~\eqref{eq:ising_diss } - \eqref{eq:ising_diss2 } represent the zero-temperature bath $\mathcal{L}_0$ \footnote{The Hamiltonian does not affect the dynamics for the simulations considered in this work and thus we set it to zero for simplicity. Physically, we interpret  the dissipative processes as a simple model for a thermalizing bath of the Hamiltonian, i.e.~we do not require dissipative engineering.
}.

\begin{figure}
    \centering
    \includegraphics[scale=0.4]{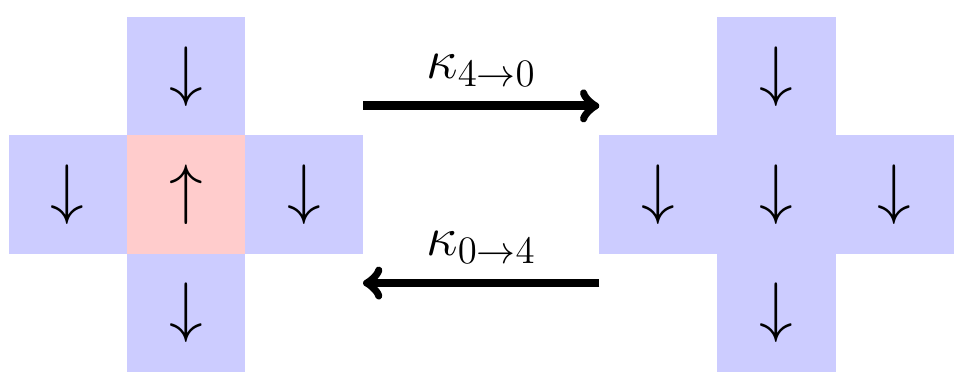}
    \caption{ The total rate of transitioning from a configuration with 4 domain walls to a configuration with 0 domain walls satisfies detailed balance: $\kappa_{0 \rightarrow 4} /\kappa_{4 \rightarrow 0 }  = e^{-8 \beta} $ where $\beta = \ln[(\kappa+\Delta_x )/ \Delta_x]/8$.  }
    \label{fig:ising_rates}
\end{figure}

If we set the rate $\tilde{\kappa} = \sqrt{ \Delta_x \kappa + \Delta_x^2} -\Delta_x$, then the steady state   is the thermal state of the 2D  Ising model:
\begin{equation} \label{eq:ising_thermal}
\rho_{ss}  =  \frac{e^{- \beta H_{is}}}{\text{Tr}[e^{- \beta H_{is}}]}, \qquad \beta = \frac{1}{8} \ln \left[\frac{\kappa + \Delta_x}{\Delta_x} \right],
\end{equation}
with the effective (inverse) temperature of the model set by the relative ratio of the correction rate to the bit-flip rate.  This is most easily understood within the quantum jump picture \cite{daley2014} since the  rates of transitioning between different classical configurations will respect detailed balance.  For example, the transition rate from a ferromagnetic configuration ($0$) to a configuration with  four domain walls ($4$)  satisfies the relationship:
\begin{equation} 
\frac{\kappa_{0 \rightarrow 4}}{\kappa_{4 \rightarrow 0}}
= \frac{\Delta_x }{ \kappa+\Delta_x} = e^{- (\Delta E) \beta} = e^{ -8 \beta}.
\end{equation} 
(See Fig.~\ref{fig:ising_rates}.) 

\subsection{Thermal steady states}\label{sec:Ising_thermal}

It is known that the 2D Ising model has a thermal phase transition in the sense that the two ferromagnetic states have an exponentially long lifetime (in $N$) when $\beta > \beta_c \approx 0.44$ \cite{kardar2007}. This is because excitations come in the form of domain walls with an energy that is proportional to their perimeter, and hence an extensive energy barrier separates the two ferromagnetic states as $N \rightarrow \infty$ \cite{eric2002, brown2016}. 

In the ferromagnetic phase ($\Delta_x \ll \kappa$) and in the limit of no dephasing ($\Delta_z=0$), the steady state of the model can support a qubit: \begin{equation} \label{eq:thermal_ising}
\rho_{ss} =  \sum_i \frac{e^{- \beta E_i}}{\mathcal{Z}}   \left(
\begin{matrix}
| E_i^+ \rangle, &
|E_i^- \rangle 
\end{matrix}
\right) 
 \left(
\begin{matrix}
|c_0|^2 & c_0 c_1 \\
 c_0^* c_1^* & |c_1|^2
\end{matrix}
\right)  \left(
\begin{matrix}
\langle E_i^+ | \\
\langle E_i^- | 
\end{matrix}
\right) ,
\end{equation}
for $|c_0|^2 + |c_1|^2=1$. $\mathcal{Z}$ is the partition function, and the states $| E_i^\pm \rangle$ are energy eigenstates of the Ising Hamiltonian which are twofold degenerate and labeled by their parity: $P | E_i^\pm \rangle = \pm | E_i^\pm \rangle$ with $P = \Pi_j X_j$.  This is a ``noiseless subsystem'' that was described in Sec.~\ref{sec:qm}. The on-diagonal degrees of freedom in Eq.~\eqref{eq:thermal_ising} are protected by a ``strong'' $\mathbb{Z}_2$ symmetry \cite{prosen2012,albert2014}: $[L_j,P]=0, \forall j$ which is generally fragile. (See Appendix \ref{sec:strong_sym} for details on the block decomposition of a Lindbladian with strong and/or weak symmetry.)

In the more physical case when both the bit flip and phase flip rate is nonzero  (i.e.~$\Delta_x \neq 0$,  $\Delta_z\neq0$) then we only get a classical bit in the low-temperature phase:
\begin{align}
\rho_{ss} &=  \sum_i \frac{e^{- \beta E_i}}{Z} \left(    \left(  | E_i^+ \rangle  \langle  E_i^+| + | E_i^- \rangle  \langle  E_i^-  | \right)/2 \right.  \\
&+ \left.  (2c-1)  \left(  | E_i^+ \rangle  \langle  E_i^-| + | E_i^- \rangle  \langle  E_i^+  | \right) /2 \right) \\
& \approx c  | \uparrow \uparrow \cdots \rangle  \langle \uparrow \uparrow \cdots | + (1-c)| \downarrow \downarrow \cdots \rangle  \langle \downarrow \downarrow \cdots |, 
\end{align}
for $c \in [0,1]$.  In this case, the system has a ``weak'' $\mathbb{Z}_2$ symmetry at the level of the full Lindbladian: $[\mathcal{L}, \mathcal{P}]=0, \mathcal{P}(\rho) = P \rho P$, and the steady states spontaneously  break this symmetry \cite{lieu2020, lieu2023b}. 

\subsection{Numerics}

\begin{figure}
    \centering
    \includegraphics[scale=0.28]{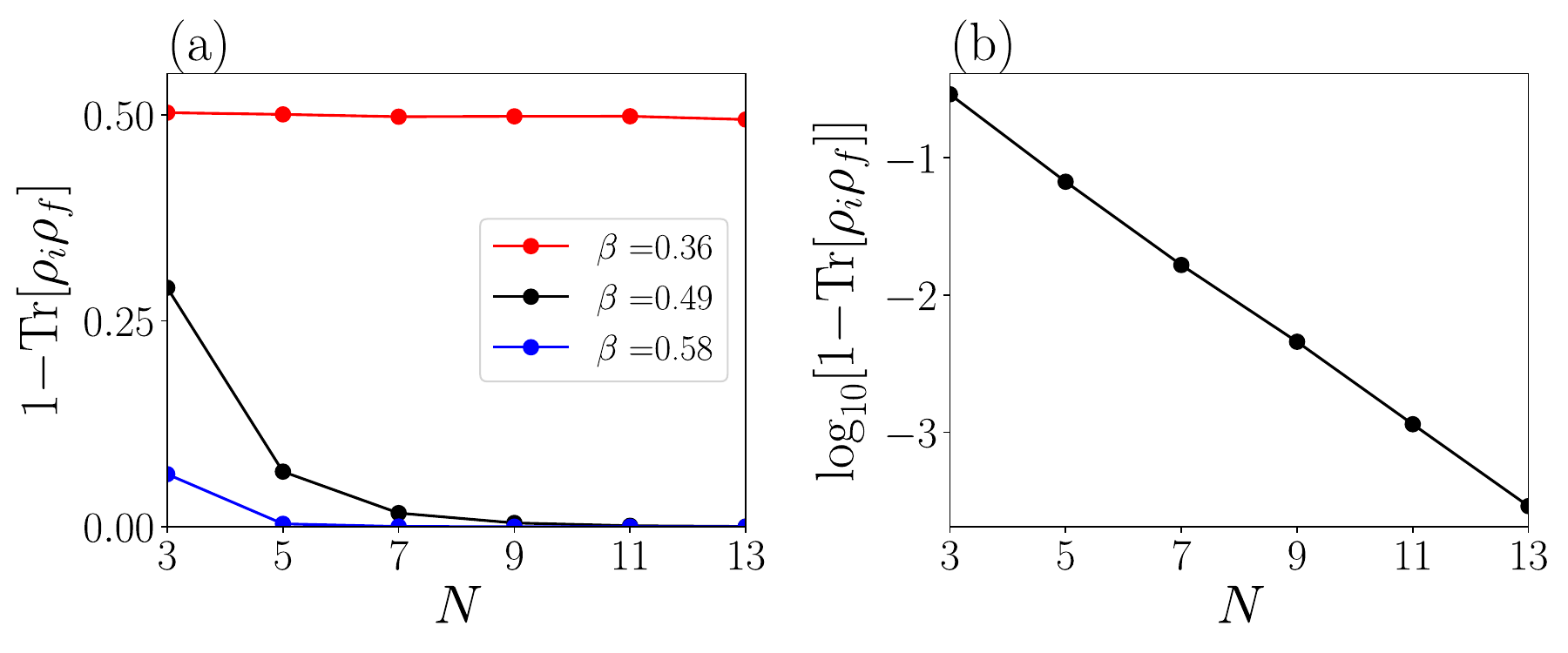} 
    
    \caption{  
    (a) The overlap between the initial and final states for the protocol given in the main text, for a Lindbladian in the high-temperature phase (red dots), and in the low-temperature phase (black and blue dots).  As linear system size $N$ grows, the overlap approaches one  only in the low-temperature (symmetry-broken) phase corresponding to $\beta > \beta_c \approx 0.44$. (b) Same black data points on a log plot; the overlap tends to one exponentially fast in $N$. In both (a) and (b), 
    the quench time is $t = 800 / \kappa $, i.e.~long enough to reach the steady state. The simulation employs the quantum jump approach by averaging over $10^5$ trajectories.}
    \label{fig:ising_error_correction}
\end{figure}

Suppose we initialize our system in a ferromagnetic state: $| \psi \rangle = |\uparrow \uparrow \uparrow\cdots \rangle $; we then quench the system with the Lindbladian described above for a time $t$ which is long enough for the system to settle into its steady state. Finally, we apply a single-shot decoder which brings the state back to the code space via a global majority rule. We apply the  following channel superoperator:
\begin{equation}
\mathcal{E} (\rho) = \sum_{\bf{r}} F_{\bf{r}} \rho F_{\bf{r}}^\dagger, \qquad F_{\bf{r}} = U_{\bf{r}} P_{\bf{r}},
\end{equation}
where $P_{\bf{r}} =  \Pi_j (1 + (-1)^{r_j} S_j) / 2$ is a projector onto a particular domain wall configuration, and $U_{\bf{r}} =  \Pi_{k \in d_{\bf{r}}} X_k$ flips all  spins $k$ in the smaller domain.  Fig.~\ref{fig:ising_error_correction} plots the overlap between the initial and final states as a function of the system size.
In the low-temperature phase, the overlap starts to approach one exponentially fast in $N$, meaning that the logical error rate drops to zero in the thermodynamic limit. Qualitatively different behavior occurs in the high-temperature phase (red dots). Here, the  overlap saturates to $0.5$ for all values of $N$.

\subsection{Connection to classical Glauber dynamics and the dissipative gap} \label{sec:metropolis}
\begin{figure}
    \centering
    \includegraphics{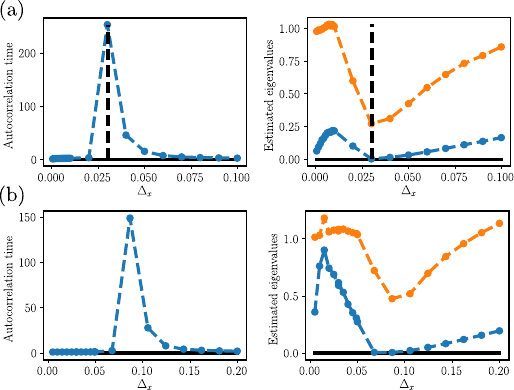}
    \caption{Estimated autocorrelation time for the dissipative Ising model on $30\times 30$ square lattice with periodic boundary condition and $\kappa = 1$. The data is obtained from 100 trajectories, each consisting of $2\times 10^5$ global time steps. The autocorrelation time and the estimate of the eigenvalues are obtained using single and two exponential fittings, respectively. The autocorrelation time $\tau$ is given in units of $\delta t = (\kappa + \Delta_x)^{-1}$. The estimated eigenvalues are given in units of $(\delta t)^{-1}$.  After fitting the function of the form $c_1e^{-\gamma_1 t}+c_2e^{-\gamma_2t}$ with $\gamma_1 \leq \gamma_2$, an estimate for the  eigenvalues of the Lindbladian is given by $\gamma_{1/2}$,
    as discussed in Sec.~\ref{sec:metropolis}.
    (a) With detailed balance, the dashed line is the critical noise rate corresponding to the critical temperature of the 2D Ising model. (b) The majority-rule case (without detailed balance).}
    \label{fig:auto_corr}
\end{figure}

We have described a set of dissipators that essentially perform  Glauber-type Monte Carlo updates on the state for every quantum jump \cite{newmanb99}. The Glauber dynamics is an efficient way to sample from the equilibrium distribution of the classical Ising model. It works as follows: First pick a spin at random, then  flip it with a probability that depends the resulting change in local energy $\Delta E$.
We can utilize results from the vast literature on Glauber dynamics to make inferences on the properties of the Lindblad spectrum. Our analysis suggests that  the Lindblad spectrum should have a nonzero dissipative gap at a generic point away from the critical point \footnote{It should be noted that the Lindbladian is gapless in the limit of zero noise ($\Delta_x=0$) since  domain walls that include an extensive area  can take a time polynomial in the linear system size to  completely shrink to a ferromagnetic configuration \cite{Verstraete:2009, Weimer2010}. However we provide evidence that  this 
gapless behavior is restricted to the fine-tuned case of zero noise.}.

It is well known that the correlation time of magnetic fluctuations diverges  at the critical temperature for Ising-Glauber (or Metropolis-Hastings) simulations \cite{newmanb99, landau1991}. Let us define the magnetic autocorrelation function:
\begin{align}
    \chi(t) &= \int dt' [m(t') - \langle{m} \rangle] [m(t'+t) - \langle{m} \rangle] \\
    &= \int dt' [m(t') m(t'+t) - \langle{m} \rangle^2] ,
\end{align}
where $m(t)$ is the time-dependent magnetization of a single spin evolving via the Glauber dynamics of the Ising model at equilibrium. Physically this measures the correlation between fluctuations in time about the average value. This function decays exponentially away from the critical point: $\chi(t) \sim e^{- t/\tau}$, where $\tau$ is the correlation time. ($\tau$ diverges at the critical point.)

Within the Lindblad formalism, the magnetic autocorrelation function can be expressed as:
\begin{equation}
    \chi(t) = \text{Tr}[  Z e^{\mathcal{L}t} ( Z \rho_{ss} ) ] - \text{Tr}[  Z  \rho_{ss}  ]^2 ,
\end{equation}
where $Z$ is the Pauli operator associated with an arbitrary spin in the lattice and $\rho_{ss} $ is the steady state. We can express this in the eigenbasis of the Lindbladian:
\begin{equation}
    e^{\mathcal{L}t} ( Z \rho_{ss} )  = \sum_{j=0} e^{\lambda_j t} c_j r_j = \text{Tr}[Z \rho_{ss}] \rho_{ss} + \sum_{j \neq 0} e^{\lambda_j t} c_j r_j ,
\end{equation}
where $l_j, r_j$ are the left and right eigenoperators of $\mathcal{L}$, $c_j = \text{Tr}[ l_j^\dagger Z \rho_{ss} ]$, and we have used the fact that the eigenoperators associated with the steady state $\lambda_0 = 0$ are $l_0 = \mathbb{I}, r_0 = \rho_{ss}$. We thus find that
\begin{equation} \label{eq:chi_rl}
    \chi(t) = \sum_{j \neq 0} e^{\lambda_j t} c_j  \text{Tr} [ Z r_j ].
\end{equation}
An autocorrelation function $\chi(t)$ that decays exponentially in time would thus be consistent with a Lindbladian that has a nonzero 
dissipative gap\footnote{We use local observables to probe the dynamical correlation, which capture the relaxation to locally stationary states and will in general not be sensitive to global properties such as the steady-state degeneracy of the system. } i.e.~$-\text{Re}[\lambda_j]>0, \forall j>0 $. (Note: Here we use $\rho_{ss}$ to mean the unique steady state in the trivial phase, and one of the symmetry-broken ferromagnetic states in the nontrivial phase; in the latter case $\text{Tr} [ Z \rho_{ss} ] \neq0$.)

To simulate this correlator, we use discrete channel evolution that is very similar to a global update of the lattice under Glauber dynamics. The number of jumps that occur during an interval of time $t$ obeys the Poisson distribution:
\begin{equation}\label{eq:poisson}
   \rho(t) = e^{\mathcal{L}t}[\rho(0)] = \sum_{k = 0}^{\infty} \Lambda^{k} [\rho(0)] \left[ \frac{\left(t\sum_s\kappa_s\right)^k}{k!} \right] e^{-t\sum_s\kappa_s} ,
\end{equation}
where the $\Lambda$ is the channel superoperator associated with a Glauber-type  single jump occurring in the system (see Appendix~\ref{sec:channel} for the derivation), and $\kappa_s$ labels all of the jump rates.  The spectrum of $\hL$ satisfies: $\text{Spec}(\hL) = \left(\sum_s\kappa_s\right)\text{Spec}(\Lambda) - \left(\sum_s\kappa_s\right)$. In a time step $\delta t = 1 / (\kappa + \Delta)$ the average number of jumps will be $[N^2(\kappa + \Delta) ] \delta t = N^2 $, i.e.~each spin on the lattice will get one update on average. If we define the channel operator of this one global update rule as $\Lambda_g = \Lambda^{N^2}$, then we will approximate  the Lindblad dynamics via the following channel evolution:
\begin{equation}
e^{\mathcal{L} (M \delta t)} \approx  \Lambda_g^{M}.
\end{equation}
The resulting dynamics  for the autocorrelator decays expotentially as a function of $t$ (away from the critical point). We numerically extract the characteristic decay time ($\tau$) and plot it as a function of error rate in the left panel of Fig.~\ref{fig:auto_corr}(a). We see that it diverges precisely at the error rate that corresponds to the critical temperature of the 2D Ising model.

We can more accurately estimate the  low-lying eigenvalues of the Lindbladian by fitting a sum of exponential functions for the value of $\chi(t)$~\cite{landau1991}. [See Eq.~\eqref{eq:chi_rl}.] We find good agreement for a sum of two exponentials for the error rates that we have scanned. (See Appendix~\ref{sec:2exp}.)  We fit the decay rate of the two exponentials, then estimate the Lindblad eigenvalues by taking their inverse. This is plotted as a function of error rate in the right panel of Fig.~\ref{fig:auto_corr}(a). The  smallest eigenvalue (estimating the gap) indeed  touches zero at the critical temperature. 

Another feature of the Glauber-like Ising dynamics   is that starting from an arbitrary initial state,  the system evolves towards the thermal distribution exponentially fast in time (away from the critical point), thus allowing to efficiently sample from the thermal stationary distribution. We can confirm that this behavior occurs in the model described above (see Appendix~\ref{sec:equilibration}), which is consistent with a finite gap in the Lindbladian \cite{kastoryano2013}. 

\subsection{Perturbing away from equilibrium}

An advantage of formulating the dynamics in terms of the Lindbladian is that we can start to perturb the system via terms that explicitly break detailed balance to test whether the stability of the phase is linked to thermal equilibrium or rather the locality and symmetry properties of the model. As a simple example, we set the two correction rates to be equal to each other: $\tilde{\kappa} = \kappa$. This corresponds to a local majority rule, i.e.~spins flip with a uniform rate $\kappa$ if the majority of neighbors are misaligned. This violates the detailed balance condition in Eq.~\eqref{eq:thermal_ising} but intuitively the error correcting properties of the phase should persist since the correction processes have a  higher rate than in the thermal case. 

In Fig.~\ref{fig:auto_corr} we show that the critical properties of the model appear to be very similar to the thermal case, i.e.~the autocorrelation time diverges at a specific value of $\Delta_x$ and the estimated Lindblad gap approaches zero at this point. One notable difference is that the critical error rate   is higher than in the thermal case, which intuitively makes sense since  the correction rate is larger.

\subsection{Effect of nonzero magnetic field}

So far we have considered errors of the form: $L \sim X,Z$, i.e.~bit flips and phase flips. In the presence of both of these errors, the Lindbladian still has a weak $\mathbb{Z}_2$ symmetry: $[\mathcal{L}, \mathcal{P}]=0, \mathcal{P}(\rho) = P \rho P$. One can ask about the stability of the classical bit with respect to perturbations that violate this condition. For example, the dissipative processes associated with a nonzero magnetic field in the $Z$ direction, i.e.~$L_j = \sqrt{\Delta_\downarrow} X_j(1-Z_j)/2$, will ensure that $[\mathcal{L}, \mathcal{P}]\neq0$. While this type of perturbation will lead to a unique steady state, the equilibration time is exponentially long. 
This dynamics has been extensively studied in the context of 2D Ising metastability \cite{binder1974}, which suggest that  the equilibration time   in the presence of a small field scales as 
$ \sim \exp[ w(\kappa, \Delta_x)/ \Delta_\downarrow]$ for some constant $w$ that depends on temperature, i.e.~the ratio of $\kappa$ and $ \Delta_x$ \cite{schonmann1994}. 

We can compare the effect of terms that explicitly break the strong and weak $\mathbb{Z}_2$ symmetry: Microscopic dephasing ($\Delta_z \neq0$) will break the strong $\mathbb{Z}_2$ symmetry, and leads to decoherence in the logical basis of the ferromagnets. The logical decoherence rate is directly proportional to the  microscopic dephasing rate $\Delta_z$. In contrast, terms that violate the weak symmetry $(\Delta_\downarrow \neq 0)$   will destroy  the logical classical bit stored in the  ferromagnets, but only at  a rate that scales as $ \sim \exp[ -  w(\kappa, \Delta_x) / \Delta_\downarrow]$. The stability of the classical bit with respect to arbitrary local perturbations is intuitively why good classical bits occur in nature (e.g.~ferromagnets and solids) even though explicit symmetry-breaking terms are always present.

\section{2D toric code}

In the previous section we suggest that the 2D Ising model is a good classical bit in the presence of generic local noise.  The rest of this work will investigate the possibility of   obtaining a topological steady state degeneracy in the Lindbladian such that arbitrary local errors will not corrupt the qubit. We will first attempt to do this in the 2D toric code \cite{kitaev2003}.

\begin{figure}
\centering
\includegraphics[scale=0.5]{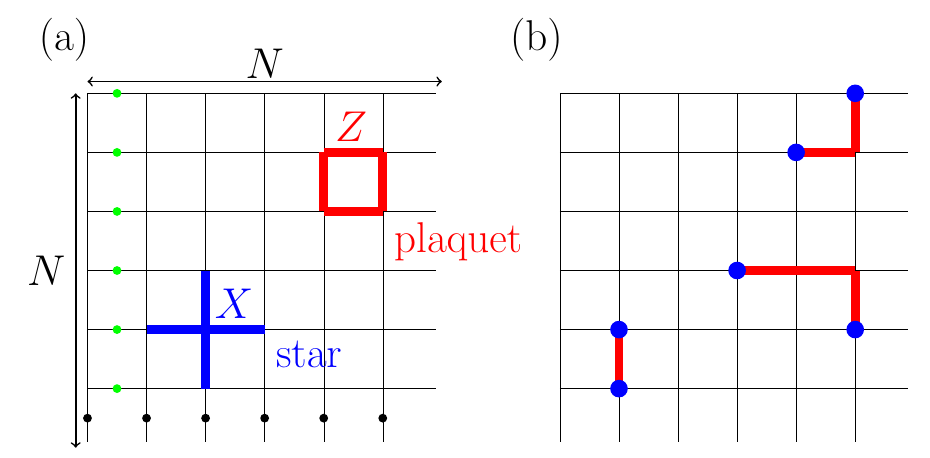} 
\caption{(a) Physical qubits live on the edges of the black squares with two qubits per unit cell. We consider an $N\times N $ lattice with periodic boundary conditions.  Star and plaquet terms couple nearest-neighbor sites. The black and green dots represent sites along $g_x$ and $g_y$ respectively. (b) Eigenstates are labeled by an $N^2$-dimensional vector $\vec{k}$ which labels the excited stars (blue dots); excited states are constructed by acting $Z$ operators  (red lines) on the ground state.}
\label{fig:tc-ham}
\end{figure}

We consider qubits that live on the edges of a square lattice with periodic boundary conditions and  $N\times N$ unit cells.  The toric code Hamiltonian reads
\begin{equation}
H_{tc} = -\sum_s A_s - \sum_p B_p, \label{HTC}
\end{equation}
where $A_s = \prod_{i\in s} X_i$ and $B_p = \prod_{i\in p}Z_i$ are stabilizers at each  vertex $s$ and plaquet $p$. (See Fig.~\ref{fig:tc-ham}.)  The ground states of the model satisfy: $ A_s  \ket{  \text{gnd} } = B_p  \ket{  \text{gnd} } =   \ket{  \text{gnd} }$, i.e.~they are eigenstates of all of the star and plaquet operators with eigenvalue $+1$.  With periodic boundary conditions,  the ground states are robustly four-fold degenerate.

The model has a gauge symmetry: $[H_{tc}, A_s] = [H_{tc}, B_p] =[A_s,B_p]=0$ which implies that eigenstates are labeled by  which star and plaquet terms are violated.  Note that star and plaquet excitations must come in pairs, i.e.~there is no way to excite a single star without exciting  another star too.

To simplify our analysis, we focus on the case where only star excitations are allowed, i.e.~none of the  plaquets are  excited. (Our main conclusions  will hold in the presence of both types of excitations.) The eigenvalues of $B_p$ are thus good quantum numbers, and we focus on the gauge sector where $B_p=+1$, i.e.~a subspace which contains the ground states of the toric code Hamiltonian.  The reduced Hilbert space will consist of states which have an even number of star excitations, and are labeled by:
\begin{align} \label{eq:convention1}
&\ket{0,0;  \vec{0} } \sim \prod_i (1+A_i) \ket{ \text{vac} }, \\
&\ket{r_x , r_y;  \vec{0} } = (g_x)^{r_x} (g_y)^{r_y}  \ket{0,0;  \vec{0} }, \\
&\ket{r_x , r_y; \vec{k} } = \left(  \prod Z \right)_{\vec{k}} \ket{r_x , r_y;, \vec{0}},
\end{align}
where $A_i$ represents different stars,  $Z_j \ket{ \text{vac} }  =  \ket{ \text{vac} }, \forall j $; $r_x , r_y; \in 0,1$ label different topological sectors;  $g_{x/y} = \Pi_{\text{hor/vert}} X$ is a product of $X$ operators along a  string (on the dual lattice) which wraps around the horizontal/vertical direction of the torus. (See Fig.~\ref{fig:tc-ham}.) The states that are labeled by $(r_x , r_y;  \vec{0})$  are orthogonal ground states of $H_{tc}$, while the  states that are labeled by  $ (r_x , r_y, \vec{k}) $ are excited states;  $\vec{k}$ is an $N^2$-dimensional vector which labels the excited stars with $1$ and non-excited stars with $0$. Excited eigenstates are defined by  applying strings of $\left(  \prod Z \right) $ operators on the ground state  via the the smallest number of $Z$ operators.

Consider the following dissipators  at each vertex (star) of the lattice: 
\begin{align}
        & L_m^{l}= \sqrt{\kappa} Z_{m,l}(1-A_m)/2, \\
        &L_m^{u} = \sqrt{\kappa} Z_{m,u}(1-A_m)/2, \\
        &L_m^{r} = \sqrt{\kappa} Z_{m,r}(1-A_m)/2,\\
        &L_m^{b} = \sqrt{\kappa} Z_{m,b}(1-A_m)/2,
\end{align}
where $l,r,t,b$ labels the left, right, top and bottom leg of the vertex at $m$. The dissipators: $L_m^{l/u/r/b}$  first check that the star $A_m$ is excited; if so, then they  will flip one of its four connecting bonds such that $A_m$ becomes de-excited (and its neighboring star stabilizer will flip).  This type of dynamics will cause the star excitations to perform a random walk on the lattice until pairs eventually meet up and annihilate each other. This model has been studied as a way to dissipatively prepare the ground state of the toric code on   a Rydberg atom simulator~\cite{Weimer2010, dengis2014}.  We also consider the effects of uniform  dephasing that acts on each physical qubit:
\begin{equation}
    L_i= \sqrt{\Delta_z} Z_i.
\end{equation}

The steady state of the Lindbladian is the thermal state of the 2D toric code: 
\begin{equation}
\rho_{ss}  =  \frac{e^{- \beta H_{tc}}}{\text{Tr}[e^{- \beta H_{tc}}]}, \qquad \beta = \frac{1}{4} \ln \left[\frac{2 \kappa + \Delta_z}{\Delta_z} \right],
\end{equation}
with the effective temperature of the model set by the relative ratio of the correction rate to the dephasing rate.   The transition rates between different stabilizer configurations  obey  detailed balance with respect to the effective  temperature $\beta^{-1}$.

\subsection{Lack of protection}\label{sec:fragile}

The 2D toric code does not have a thermal phase transition, i.e.~the critical properties are strictly a zero-temperature effect \cite{brown2016, Alicki2009, eric2002}. (This is analogous to lack of thermal stability of the quantum phase transition in the 1D  Ising model.) Intuitively,  this is because there is no extensive energy barrier  between degenerate ground states, i.e.~one ground state can evolve to another via a single anyonic string excitation that costs a constant amount of energy.

Let us  describe the thermal steady state of this model.  Within the $B_p=+1$ gauge sector,  we can   partition the  subspace into different topological sectors. We define the projection operators:
\begin{equation} \label{eq:projectors}
P_{ r_x, r_y} = \sum_{ \vec{k} } \ket{r_x, r_y; \vec{k} } \bra{ r_x, r_y;  \vec{k} } , 
\end{equation}
where $r_x, r_y \in 0,1$; $P_{r_x, r_y}$ projects states into topological sector $r_x, r_y$\footnote{One can construct global string operators similar to $g_{x/y}$, which consists of a product of $Z$ operators along a vertical or horizontal string on the lattice. The projector $P_{ r_x, r_y}$ projects onto the eigenstates of the $Z$ global string operators.}.  Note that all of the  dissipators  will commute with $P_{r_x, r_y}$, thus these projectors are strong symmetries of the Lindbladian \cite{prosen2012}. There exists a basis where the Lindbladian can be block diagonalized into $4^2=16$ different sectors: 
\begin{equation} \label{eq:blockLind}
    \hL= \text{Diag}[  \hL_{0,0}, \hL_{0,1}, \hL_{0,2}, \ldots \hL_{3,3} ]
\end{equation}
where the numbers 0 to 3 label four different topological sectors of the bras and kets accoding to the convention: $(r_x=0, r_y=0) \rightarrow 0, (r_x=1,  r_y=0) \rightarrow 1, (r_x=0,  r_y=1) \rightarrow 2, (r_x=1,  r_y=1) \rightarrow 3 $.  The Lindbladian $\hL_{0,0}$ acts on operators where both ket and bra belong to the same topological sector $ r_x=0,  r_y=0$. $\hL_{0,1}$ acts on operators where the ket belongs to sector  $ r_x=0,  r_y=0$, while the bra belongs to sector $ r_x=1,  r_y=0$.

The effect of noise on the off-diagonal sectors such as $\hL_{0,1}$ can be estimated using an argument based on anyon random walk~\cite{chesi2010, brown2016}. To the leading order in the perturbation, the noise creates a single pair of exitations above the steady state. After some time, the exitations will either annihilate with their partner locally or they get separated by a distance $N/2$ apart, leading to a global loop operator that decoheres the state. The  probability that a 2D square lattice random walker does not return to its initial position after $L$ steps scales like $1/\ln(L)$. We can estimate that the probability of decoherence scales as $\Delta_z/\ln(N/2)$. We therefore expect that for relatively small system size, the off-diagonal sectors are gapped by the noise with an eigenvalue $O(\Delta_zN^2/\ln(N/2))$ for small $\Delta_z$. Indeed, a more careful analysis for generic $N$ reveals that the gap scales with $O(\Delta_z)$~\cite{Alicki2009,chesi2010,brown2016}.

We will provide numerical evidence in the next section that the steady state structure will be the following for any non-zero temperature:
\begin{equation}
\rho_{ss} = \sum_i \frac{e^{-\beta E_i}}{ \mathcal{Z}} \left( \sum_{ r_x, r_y = 0}^1 c_{ r_x, r_y} |  r_x, r_y; E_i \rangle \langle   r_x, r_y ; E_i |  \right)
\end{equation}
where $E_i$ labels the energy, $|  r_x, r_y; E_i\rangle$ is the corresponding eigenstate in topological sector $r_x,r_y$, and  $\sum_{ r_x, r_y} c_{ r_x, r_y}=1$.  We therefore find that coherences between different topological sectors are not stable. This implies that only a classical bit can be stored in the steady state. We also note that this classical bit structure is an artifact of imposing the gauge symmetry, i.e.~the presence of bit flips ($L \sim X$)   will remove all strong symmetries, thus reducing the  classical bit  to a unique thermal steady state. 

\subsection{Numerics}

\begin{figure}
    \centering
    \includegraphics[scale=0.35]{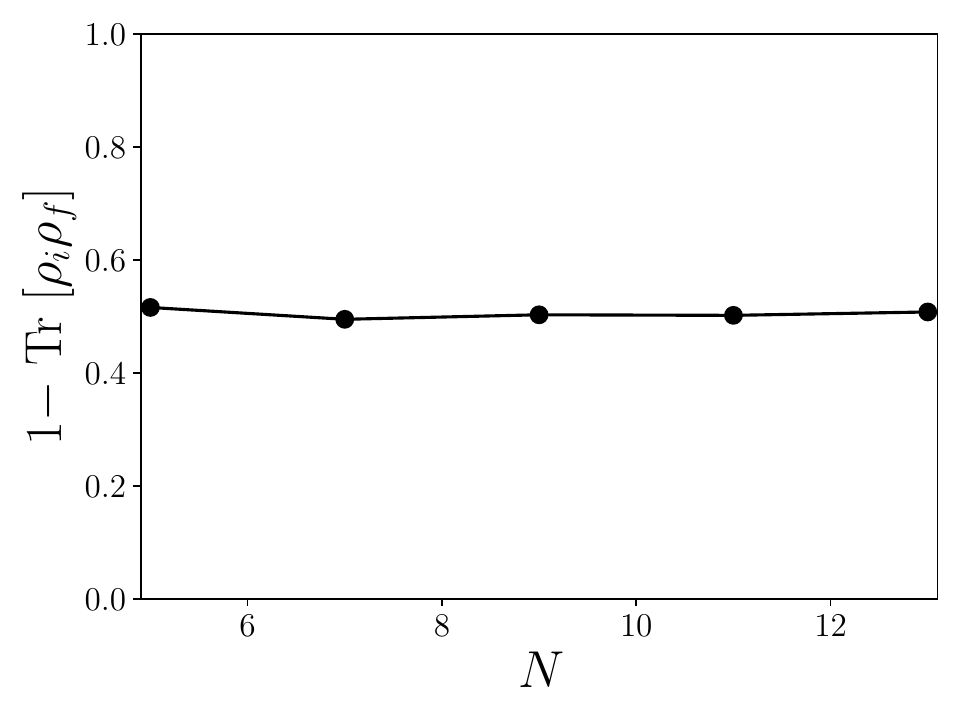}
    \caption{  The overlap between the initial and final states for the protocol given in the main text with  with $\Delta_z / \kappa= 0.01$ ($\beta = 1.3$). Unlike the Ising model, there is no ``threshold'' behavior, i.e.~any non-zero temperature causes coherences to decay.  $t = 20/\kappa$ and we  average over $10^3$ trajectories.}
    \label{fig:tc_error_correction}
\end{figure}

Suppose we initialize our system in a superposition of ground states in different topological sectors: $| \psi \rangle = (| 0,0; E_0 \rangle + | 1,0; E_0 \rangle ) /\sqrt{2} $. We then quench the system with the Lindbladian described above for a time $t$ which is long enough for the system to settle into its steady state. Finally, we apply a single-shot decoder which brings the state back to the code space via the channel superoperator:
\begin{equation}
\mathcal{E} (\rho) = \sum_{\bf{r}} F_{\bf{r}} \rho F_{\bf{r}}^\dagger, \qquad F_{\bf{r}} = U_{\bf{r}} P_{\bf{r}}
\end{equation}
$P_{\bf{r}} =  \Pi_j (1 + (-1)^{r_j} A_j) / 2$ is a projector onto a particular  configuration of excited anyons (stars), and $U_{\bf{r}} $  is a minimal-weight matching unitary operator which sends the state back to the code space by applying a minimal number of $Z$ operators which de-excite all anyons. 
(See Fig.~\ref{fig:tc-ham}.) The initial and final state overlap is plotted in Fig.~\ref{fig:tc_error_correction}.  We find that the overlap saturates to a value of $0.5$, which suggests that any non-zero dephasing is enough to destroy coherences between ground states. There is no critical temperature below which coherences are preserved in the thermodynamic limit (apart from exactly at $\beta = \infty$ when there is no dephasing).

\begin{figure}
    \centering
     \includegraphics[scale=0.28]{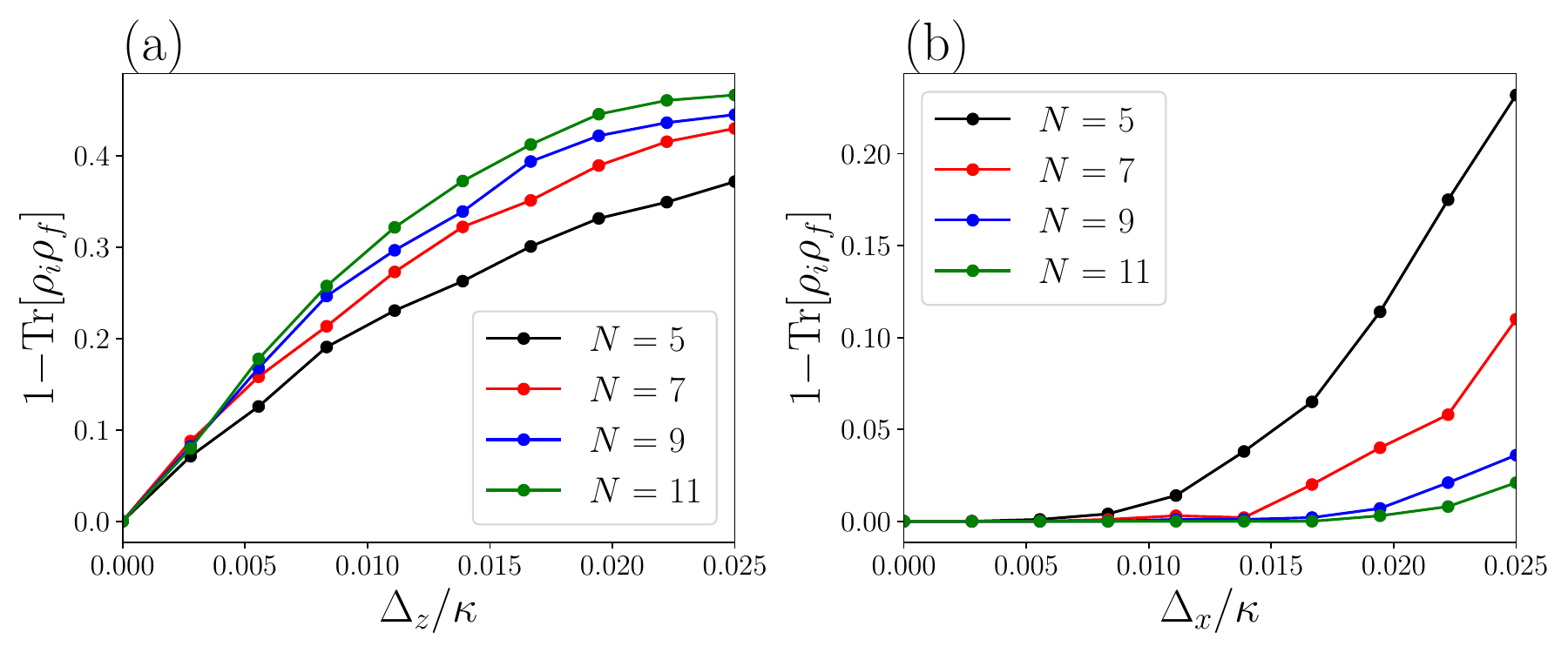} 
    \caption{ (a) 2D toric code overlap as a function of dephasing rate for a fixed quench time $ t = 3/\kappa$. For a fixed error rate and quench time, the logical error rate does not improve with system size. (b) 2D Ising model  overlap for a fixed quench time $t = 200 / \kappa $.  For a fixed error rate and quench time, the logical error rate improves with system size in the symmetry-broken phase.  Plots are averaged over $10^4$ trajectories.}
    \label{fig:ising_tc_compare}
\end{figure}

We can also notice the difference between the 2D toric code and the 2D Ising model by varying the noise rate $\Delta$
for a fixed (but finite) noise time $t$, then applying the decoder $\mathcal{E}$.   This is shown in Fig.~\ref{fig:ising_tc_compare}.  For the Ising model, we find that the logical error rate gets suppressed as we increase the system size. This is not true for the 2D toric code.

\section{4D toric code} \label{sec:4d_tc}
\begin{figure}[t]
    \centering
    \includegraphics{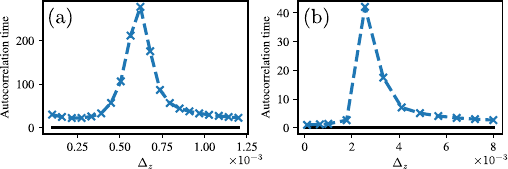}
    \caption{Extracted autocorrelation time of the mean stabilizer $\bar{S}_e$ for 4D toric code on a $5\times 5\times 5\times 5$ lattice with periodic boundary condition and $\kappa = 1$. The lattice contains 3750 spins and the data is collected from 100 trajectories, each containing $10^5$ global time steps. (a) With detailed balance, a critical point is close to $\Delta_z \approx 0.0006$. 
    (b) With the majority-vote rule, a critical point is close to $\Delta_z \approx 0.0026$.}
    \label{fig:auto_4d}
\end{figure}
We have studied a local dissipative model that prepares the thermal state of the 2D toric code, and argued that in the presence of bit flips and phase flips the model has a unique thermal steady state. We will now construct a similar model for the 4D toric code, and  suggest  that it is stable against both bit flips and phase flips.

The 4D toric code can be understood as the hypergraph product of two 2D Ising models \cite{breuckmann2021}; intuitively, one of the Ising models protects against bit flips while the other protects against phase flips. We describe salient features of the model, following the  description found in Ref.~\cite{pastawski2011}.  For every vertex of an $N \times N \times N \times N$ lattice,  one can associate 4 edges, 6 faces, and 4 cubes.  (For a 3D lattice, every vertex has 3 edges, 3 faces, and 1 cube.) Physical qubits live on each face of the lattice,  so there are $6N^4$ total physical qubits. There are two types of stabilizers  $S_e, S_c$ which are associated with the edge and cube degrees of freedom respectively.  Each physical qubit  appears in four of the $S_e$ stabilizers and four of the $S_c$ stabilizers (similar to the 2D Ising model). In the 4D toric code, unsatisfied  edge  and cube stabilizers must form a closed domain wall, which is 
ultimately responsible for the thermal stability  of the 4D toric code.

Since it is difficult to gain intuition in 4D space, it  is useful to reformulate things in a more algebraic way. Each vertex of the lattice can be associated with a four-component vector: $\vec{v} = [v_0, v_1, v_2, v_3]$ where $v_i \in[1,N]$. The edges, faces, and cubes corresponding to a particular vertex are associated with a four-component binary vector:
\begin{equation}
\hat{e}, \hat{f}, \hat{c} \in \{ (x_0,x_1,x_2,x_3) | x_i \in [0,1] \}  , 
\end{equation}
with edges $\hat{e}$, faces $\hat{f}$, and cubes $\hat{c}$ satisfying the condition $\sum_i x_i $ equal to $1,2,3$ respectively.  In other words, a face is defined by two edges, and a cube is defined by three edges. There are indeed 4 edges, 6 faces, and 4 cubes per vertex. Each physical qubit is identified with a tuple $\vec{v}, \hat{f}$ which specifies both the face orientation $\hat{f}$ and the vertex $\vec{v}$.

The stabilizers associated with  edges and cubes of the lattice  are:
\begin{align}
S_{\vec{v},\hat{e}} &= \bigotimes_{\hat{e} \subset \hat{f}} X_{\vec{v}, \hat{f}} \otimes X_{\vec{v} -\hat{f} + \hat{e} , \hat{f}}, \\
S_{\vec{v},\hat{c}} &= \bigotimes_{\hat{f} \subset \hat{c}} Z_{\vec{v}, \hat{f}} \otimes Z_{\vec{v} +\hat{c} - \hat{f} , \hat{f}}.
\end{align}
Each edge $\hat{e}$ appears in 3 faces $\hat{f}$, and each face appears in 3 cubes $\hat{c}$, hence both stabilizers are a product of 6 Pauli operators. Note that a particular operator $X_{\vec{v}', \hat{f}'}$ appears in four different stabilizers $S_{\vec{v},\hat{e}}$ and that the stabilizers  commute with each other. The 4D toric code Hamiltonian reads:
\begin{equation}
H_{4d} =   - \sum_{\vec{v}, \hat{e}}   S_{\vec{v},\hat{e}} - \sum_{\vec{v}, \hat{c}}   S_{\vec{v},\hat{c}} .
\end{equation}

As before, for simplicity we restrict ourselves to only $Z$-dephasing errors which cause excitations of the $S_{\vec{v},\hat{e}}$ stabilizers. We  work in the gauge sector where all cube stabilizers are satisfied: $S_{\vec{v},\hat{c}} = +1$.  The following states span this subspace:
\begin{align}\label{eq:convention2}
    &\ket{\vec{0};  \vec{0} } \sim \prod_i (1+S_{e,i}) \ket{ \text{vac} }, \\
    &\ket{\vec{r};  \vec{0} } = \Pi_j g_{\hat{f}_j}^{r_j}  \ket{\vec{0};  \vec{0} }, \\
    &\ket{\vec{r}, \vec{k} } = \left(  \prod Z \right)_{\vec{k}} \ket{\vec{r}, \vec{0}},
\end{align}
where the product on $i$ runs over all edge stabilizers. The vector $\vec{r}$ has six components that are either $0$ or $1$. We define 6 logical operators $g_{\hat{f}_j}$, one per each face direction. They read:
\begin{equation}
g_{\hat{f}_j} = \bigotimes_{n,m=1}^N X_{n\hat{e}_3 + m\hat{e}_4, \hat{f}_j },
\end{equation}
where $\hat{e}_3 +\hat{e}_4 = (1,1,1,1) - \hat{f}_j $. These operators  commute with the stabilizers and relate states that belong to  the $2^6 = 64$ different topological sectors.

We now describe thermal dissipators of the 4D toric code that are analogous to the dissipators of the 2D Ising model. They read:
\begin{align} \label{eq:4d_diss0}
L_{v',f'}^{(4)} &= \sqrt{\kappa} Z_{v', f'} P_{v',e_1'}^{-} P_{v',e_2'}^{-} P_{v'+f'-e_1',e_1'}^{-} P_{v'+f'-e_{2}',e_2'}^{-}  \\
L_{v',f'}^{(3)} &= \sqrt{ \tilde{\kappa}}   Z_{v', f'} P_{v',e_1'}^{+} P_{v',e_2'}^{-} P_{v'+f'-e_1',e_1'}^{-} P_{v'+f'-e_{2}',e_2'}^{-}  \\
L_{v',f'}^{(2)} &= \sqrt{  \kappa}   Z_{v', f'} P_{v',e_1'}^{+} P_{v',e_2'}^{+} P_{v'+f'-e_1',e_1'}^{-} P_{v'+f'-e_{2}',e_2'}^{-} \label{eq:4d_diss},
\end{align}
where $P_{v, e}^{\pm} = (1 \pm  S_{v,e}) / 2$, and we have used the convention $e_1' + e_2' = f'$. We also consider the  presence of $Z$ dephasing on each face: $L' = \sqrt{\Delta_z} Z_{v, f}$.  For $\tilde{\kappa} = \sqrt{ \Delta_z \kappa + \Delta_z^2} -\Delta_z$ the steady state is the thermal state of the 4D toric code:
\begin{equation}\label{eq:4d_tempc}
\rho_{ss}  =  \frac{e^{- \beta H_{4d}}}{\text{Tr}[e^{- \beta H_{4d}}]}, \qquad \beta = \frac{1}{8} \ln \left[\frac{\kappa + \Delta_z}{\Delta_z} \right].
\end{equation}

\subsection{Steady-state qubit}

Constructing a single-shot decoder for the 4D toric code is more challenging than for the 2D models studied in this work. 
(Ref.~\cite{breuckmann2021} provides a description of a local decoder for the 4D toric code, but such schemes can get ``stuck'' in sheet-like configurations that are not in the code space.) It has analytically been shown that the 4D toric code is capable of storing quantum information in its thermal state in the low-temperature phase \cite{eric2002,Alicki2010}, and therefore we expect that the steady state of our local Lindblad model  at low-temperature will assume the form
\begin{equation} \label{eq:noiseless_sub}
\rho_{ss} =  \sum_i \frac{e^{- \beta E_i}}{ \mathcal{Z}}  \left(
\begin{matrix}
| E_i^{\vec{r}} \rangle, &
|E_i^{\vec{s}}  \rangle 
\end{matrix}
\right) 
 \left(
\begin{matrix}
|c_0|^2 & c_0 c_1 \\
 c_0^* c_1^* & |c_1|^2
\end{matrix}
\right)  \left(
\begin{matrix}
\langle E_i^{\vec{r}}  | \\
\langle E_i^{\vec{s}}  | 
\end{matrix}
\right) ,
\end{equation}
for $|c_0|^2 + |c_1|^2=1$, i.e.~coherences and populations between different topological sector are protected. Similar thermal dissipators to the ones in Eqs.~\eqref{eq:4d_diss0}-\eqref{eq:4d_diss} can be constructed to protect against logical bit flips $L \sim X$\footnote{Analogous to the 2D toric code (Sec.~\ref{sec:fragile}), when only one type of noise, say $Z$, is present, the Lindbladian has a strong symmetry with respect to the symmetry generated by the projectors $P_{\vec{r}} = \sum_{\vec{k}}\ketbra{\vec{r}, \vec{k} }{\vec{r}, \vec{k} }$. The emergence of the noiseless subsystem can be viewed as a result of spontaneous symmetry breaking (see Sec.~\ref{sec:Ising_thermal} and Appendix~\ref{sec:strong_sym}). Unlike $\mathbb{Z}_2$ symmetry breaking, there do not exist any local order parameters that can distinguish different symmetry-broken sectors.}. 
Since the zero-temperature bath superoperator responsible for protection against $X$ commutes with the corresponding superoperator that protects against $Z$, we expect the noiseless subsystem to protect against both sources of noise. 
We also note that dynamical simulations using a ``Toom's rule'' model that is very  similar to our local Lindbladian have demonstrated an exponential protection against both local bit flips and phase flips \cite{pastawski2011}, again corroborating the description above.  

We can  observe  signatures of the transition by considering the autocorrelation function for the stabilizers of the model. We define the mean stabilizer autocorrelation function as:  
\begin{equation}
    \chi(t) = \text{Tr}[  \bar{S}_e  e^{\mathcal{L}t} ( \bar{S}_e  \rho_{ss} ) ] - \text{Tr}[ \bar{S}_e   \rho_{ss}  ]^2 ,
\end{equation}
where $  \bar{S}_e = (\sum_{\vec{v}, \hat{e}}   S_{\vec{v},\hat{e}})/4N^4$ is the average of the edge stabilizers on the lattice\footnote{We choose the autocorrelator of the average  edge stabilizers rather than a single stabilizer since the numerical peak at the critical point is sharper for the former.}. In analogy with the 2D Ising model, we expect this correlator to decay exponentially in time away from the critical point. In Fig.~\ref{fig:auto_4d} we plot the extracted correlation time as a function of the error rate for both (a) the case of a thermalizing Lindbladian, and (b) the case of the majority rule Lindbladian ($\tilde{\kappa} = \kappa$). We find that indeed both models exhibit a diverging correlation time at a critical error rate. This is consistent with  a low-temperature regime that passively protects a qubit.

\section{Discussion and outlook}

Most studies in the field of passive quantum error correction identify dissipative processes that only provide first (or $n$th) order protection against noise. Such schemes will  require some form of active error correction (i.e.~syndrome measurements) to eventually  reach fault tolerance.  In this work, we have focused on identifying  local Lindbladians for spin systems that can  \textit{exponentially} protect against local noise. We suggest that such models are  associated with nontrivial states of matter, since the latter are characterized by robust degeneracies in the steady state of the Lindbladian. $\mathbb{Z}_2$ symmetry breaking appears useful to protect a classical bit, while intrinsic topological order protects a qubit.

An important  question is whether a Lindbladian can host a phase with intrinsic topological order in less than 4D. The area of driven-dissipative phase transitions might provide a route which has hitherto been unexplored (see Appendix~\ref{sec:driven-diss}). It should be noted that many aspects of symmetry-breaking driven-dissipative phase transitions closely resemble their thermal counterparts (e.g.~universality classes and lower critical dimensions) \cite{maghrebi2016, Sieberer_2016}, so it is unclear whether adding a quantum drive can  produce a topological transition in less than 4D. Nevertheless, this is a direction that warrants further attention.

While we have focused on topological steady state degeneracy as a good indicator of topological order in  open quantum systems,
it would be interesting to see how this compares with other recent efforts to define topological order in a mixed state \cite{hastings:2011,  grover2020, ryu2022, bao2023b, zemin2022, lee2023c, fan2023, hsieh2023}. 

It is known that  quantum phase transitions come in yet another flavor: symmetry-protected topological (SPT) phase transitions. Notable examples include topological insulators \cite{kane2010, zhang2011} and the Haldane phase of spin chains \cite{Affleck_1989,pollmann2010}. Various generalizations of open (Lindblad) SPT  matter have recently been put forward \cite{lieuTFW, roberts_2017, diehl2020,
roberts_2020, 
ueda2020,bergholtz2021,
 zhou2021,
altland2021, deGroot2022,  lee2023b,lieu2022, sato2023, cooper2023}. However, none of these studies have found robust zero-decay-rate edge modes in the Lindbladian that survive the presence of local (symmetric) perturbations. This may be because standard SPT phases are not thermally stable. Recent work \cite{roberts_2017, roberts_2020, stahl2021} has suggested that 1-form symmetries must be imposed to obtain a thermally-stable SPT phase. Perhaps such systems host a protected classical bit in the presence of both bit and phase flips, in analogy with the 2D Ising model.

Is 4D necessary to obtain a passive  quantum memory? In a recent work \cite{lieu2023b}, we have suggested that it is possible to achieve such a model in 2D by creating an Ising model out of \textit{bosonic} cat qubits, i.e.~going beyond the two-level-system approximation. (An Ising interaction can be generated by placing a Josephson junction between cavities, see SM~5 in \cite{lieu2023b}.) The driven-dissipative cat code \cite{mirrahimi2014} is a bosonic qubit that  spontaneously breaks $\mathbb{Z}_2$ photon parity symmetry and satisfies the definitions of a phase as outlined in Sec.~\ref{sec:qm} \cite{ciuti2018, lieu2020}. This is another example of a passive classical bit that is encoded in the coherent states: $|\pm \alpha \rangle$.  The 2D Ising-cat model \cite{lieu2023b} thus breaks two separate $\mathbb{Z}_2$ symmetries (i.e.~a photon parity symmetry within each cavity and an Ising parity symmetry of the lattice), one of which protects against bit flips and another which protects against phase flips. 
An interesting open question remains to find other bosonic lattice systems that have this property, and to construct experimental proposals  to realize this model on current hardware platforms.

\textit{Acknowledgements.---}We sincerely thank Victor Albert, Alexey Gorshkov, and Oles Shtanko for useful discussions. Y.-J.L acknowledges support from the Max Planck Gesellschaft (MPG) through the International Max Planck Research School for Quantum Science and Technology (IMPRS-QST) and the Munich Quantum Valley, which is supported by the Bavarian state government with funds from the Hightech Agenda Bayern Plus. S.L.~was supported by the NIST NRC Research Postdoctoral Associateship.

\textit{Note added.---}We note an independent work \cite{wang2023d} that comes to similar conclusions regarding the 2D toric code, and studies a 3D toric code model with classical steady state degeneracy.

\appendix

\section{Qubit steady state in a finite system} \label{sec:finite}

In this paper, we focus on qubit steady state structures that only emerge in the thermodynamic limit of a nontrivial phase. Here, we study a finite system that hosts a qubit steady state. We show that it does not passively protect against  local $X$ or $Z$ errors, and expect this behavior to be generic. 

Consider a Hilbert space of two qubits. We consider a single jump operator: $L = X_2(1 - Z_1 Z_2)/2$. This can be viewed as a single correction dissipator for the Ising model in the main text. The model has a qubit steady state structure: Any state of the form: $|\psi \rangle = c_0 |\uparrow \uparrow \rangle + c_1 |\downarrow \downarrow \rangle  $ is a steady state of the model. This qubit is protected by a non-Abelian strong symmetry \cite{Zhang_2020}: $[L, U_1] = [L, U_2] =0 $ where $U_1 = Z_1, U_2 = X_1 X_2 , [U_1, U_2] \neq 0$. This system is \textit{not} protected against noise in either basis, i.e.~jump operators of the form $L \sim Z_1, Z_2, X_1$ will each cause the qubit to decohere since these jumps do   not commute with both symmetries. (Note that for the Ising model in the main text, the qubit is protected against all local $X$ errors in the thermodynamic limit.)  
 
More generally (to our best knowledge), finite systems require  both logical $\bar{X}$ and $\bar{Z}$ operators ($U_1, U_2$ above) to commute with the dissipation operators in order to have a qubit steady state. Generic  noise in the $X,Z$ basis will necessarily anticommute with one of the logical operators leading to destruction of the noiseless subsystem.

In other words, to get a qubit steady state in a finite system requires us to impose \text{two} strong symmetry constraints on the noise. For $\mathbb{Z}_2$ strong symmetry breaking, we only need to impose one strong symmetry constraint to obtain a qubit steady state (in the thermodynamic limit). For intrinsic topological order, we do not have any symmetry constraints for a qubit. 

\section{Spontaneous symmetry breaking in a Lindbladian} \label{sec:strong_sym}

We briefly review the symmetry structure in a Lindbladian in the presence of ``strong'' and ``weak'' symmetries \cite{prosen2012} and the steady state solutions in a symmetry-broken phase \cite{lieu2022}, focusing on the case of $\mathbb{Z}_2$. A Lindbladian is said to have a strong $\mathbb{Z}_2$ symmetry if $[\hL, \mathcal{P}_l] = [\hL, \mathcal{P}_r] = 0 $ where $\mathcal{P}_l(\rho) = P \rho, \mathcal{P}_r(\rho) =  \rho P$ are superoperators that act on the left and right of an operator, and $P$ is a parity operator: $P^2 = \mathbb{I}$. If all of the microscopic  dissipators  of a Lindbladian commute with the parity operator: $[L_j,P]=0, \forall j$, then $\hL$ will have a strong symmetry. In this case the Lindbladian can be block diagonalized into four symmetry sectors
\begin{equation}
    \hL = \text{Diag}[\hL_{++}, \hL_{--}, \hL_{+-}, \hL_{-+}].
\end{equation}
Each sector acts on operators that  are eigenoperators of $\mathcal{P}_l$ and $\mathcal{P}_r$, with eigenvalue $\pm 1$. The sectors $\hL_{++} $  and $ \hL_{--}$ contain operators with nonzero trace, and therefore those sectors must each have an exact eigenvalue of zero, corresponding to a steady state. In a symmetry-broken phase, the off-diagonal sectors $\hL_{+-}$ and $ \hL_{-+}$ also acquire an eigenvalue of zero but only in the thermodynamic limit.  This leads to enough degrees of freedom to store a qubit in the steady state, i.e.~a noiseless subsystem. 

A Lindbladian is said to have a weak $\mathbb{Z}_2$ symmetry if $[\hL, \mathcal{P}] = 0 $ where $\mathcal{P}(\rho) = P \rho P$ is a parity superoperator that acts on bras and kets simultaneously. 
Physically, this means that the symmetry $P$ is conserved when the system and its environment are both taken into account.
This expression can be satisfied even if some of the dissipators anticommute with the parity, and hence it is a weaker condition. In this case the Lindbladian can be block diagonalized into two symmetry sectors
\begin{equation}
    \hL = \text{Diag}[\hL_{+}, \hL_{-}].
\end{equation}
Each sector acts on operators that  are eigenoperators of $\mathcal{P}$ with eigenvalue $\pm 1$. Only $\hL_{+}$ acts on traceful operators, hence a weak symmetry by itself do not imply multiple steady states. However, in a symmetry-broken phase, the off-diagonal sectors $\hL_{-}$ also acquires an eigenvalue of zero in the thermodynamic limit.  This leads to enough degrees of freedom to store a classical bit in the steady state.

\section{Unraveling the dynamics of the Lindbladian} \label{sec:channel}
In this section we  describe some generic features of the Lindblad models considered in the main text.
Denote by $S$ the set of states $\ketbra{\phi}{\phi'}\in \mathcal{H}\otimes\mathcal{H}$ where $\ket{\phi}$ and $\ket{\phi'}$ are eigenstates with the same eigenvalue for  all of the stabilizers.
Note that this does not imply $\ket{\phi} = \ket{\phi'}$. We also let $D$ be the set  of states $\ketbra{\phi}{\phi'}\in \mathcal{H}\otimes\mathcal{H}$ where $\ket{\phi}$ and $\ket{\phi'}$ differ by at least one stabilizer value. The subspaces spanned by $S$ and $D$ form a bipartition of the entire Hilbert space $\text{Span}(S)\oplus \text{Span}(D) = \mathcal{H}\otimes\mathcal{H}$.
It is useful to notice that in all the stabilizer models we considered in the main text, the dynamics are decoupled between the subspaces spanned by $S$ and $D$. More precisely, if $A\in \text{Span}(S)$, then $e^{\hL t}A\in  \text{Span}(S)$. The same holds for the set $D$.

We will now show that the Lindbladians we considered in the main text are generically gapped within the subspace $\text{Span}(D)$.
Suppose $A\in D$. Recall that the Lindbladians in the main text take a form
\begin{equation}
    \hL(\rho) = \sum_s\hL_s = \sum_s\kappa_s\left(L_s\rho L_s^{\dag}-\frac{1}{2}\{L^{\dag}_sL_s,\rho\}\right),
\end{equation}
for some dissipative rates $\kappa_s\geq 0$. The protection part has jump operators of the form $L_s = U_sP_s$, where $U_s$ is some Pauli operator and $P_s$ is a projector onto some particular local stabilizer configuration. Let $\hL_A = \sum_{s\in C_A} \hL_s$, where $C_A$ is the set of indices for the terms of the protection part for which the stabilizer values in $A$ mismatch in its bra and ket. Apply $\hL_A$ to $A$ we find that the terms $L_sAL_s^{\dag}$ vanish due to the mismatch of stabilizer values, only the terms $\{L_s^{\dag}L_s, A\} = P_sA + AP_s\propto A$ contribute. Therefore, $A$ is a right eigenvector of $\hL_A$ with a negative eigenvalue. Since $\hL= \hL_A + \sum_{s\notin C_A}\hL_s$ and $\sum_{s\notin C_A}\hL_s$ is itself a Lindbladian whose eigenvalues must have a non-positive real part. It follows that $\hL$ must have a gap greater than the gap of $\hL_A$. So $\hL$ is gapped in $\text{Span}(D)$.

One may notice that there exist highly fine-tuned cases where $ P_sA + AP_s = 0$ for all $s\in C_A$. This can happen, for instance, when the domain walls in 2D Ising model or the 4D toric code contain no corners and are straight across the entire system. However, we expect these configurations to be unstable under any non-zero noise, and they will be rapidly destabilized into a mixture consisting of mostly non-fine-tuned configurations.

Next, we will show that the Poissonian unraveling Eq.~\eqref{eq:poisson} is valid in $\text{Span}(S)$. Therefore, the autocorrelation extracted in the main text is relevant for the spectrum of the Lindbladian in $\text{Span}(S)$. Consider $A\in \text{Span}(S)$, then any Lindbladian with jump $L_s = P_s$ and $P_s$ being a projector onto some local stabilizer configuration will annihilate $A$. Within the subspace $\text{Span}(S)$, inserting these ``do-nothing'' jumps does not change the dynamics. By adding appropriately chosen do-nothing jumps, the Lindbladians in the main text can be made to satisfy $\sum_s \kappa_sL^{\dag}_sL_s =\sum_s\kappa_s$. In this case we can define a completely-positive-trace-preserving   map $\Lambda(\rho) = \sum_s \kappa_s L_s\rho L_s^{\dag}/(\sum_s\kappa_s)$ such that
\begin{equation}
    \hL(\rho) = \left(\sum_s\kappa_s\right)\left(\Lambda(\rho)-\rho\right).
\end{equation}
By Taylor expanding the time-evolution operator $e^{\hL t}$ using this relation, we obtain the relation Eq.~\eqref{eq:poisson} in the main text. 

For the 2D Ising model and the 4D toric code, we have $(\sum_s\kappa_s) = n(\kappa+\Delta)$, where the rates $\kappa$ and $\Delta$ ($\Delta_z$ or $\Delta_x$) are the same as defined in the main text, and $n$ denotes the number of physical qubits in the system.
The channel operator takes the form
\begin{equation}
    \Lambda(\rho) = \frac{1}{\kappa + \Delta}\left(\kappa\Lambda_r(\rho)+\Delta\Lambda_e(\rho)\right),
\end{equation}
 The noise channel $\Lambda_e$ is given by
\begin{equation}
    \Lambda_e(\rho) = \frac{1}{n}\sum_i Z_i\rho Z_i\ \text{ or }\ \Lambda_e(\rho) =\frac{1}{n}\sum_i X_i\rho X_i,
\end{equation}
where the index $i$ sums over all the physical qubits.
The protecting channel $\Lambda_r$ is given by
\begin{equation}
    \Lambda_r(\rho) = \frac{1}{n} \sum_i \sum_m \left(L^{(m)}_i\rho  (L^{(m)}_i)^{\dag} + \gamma_m P^{(m)}_i\rho  (P^{(m)}_i)^{\dag}\right)
\end{equation}
where the index $m$ sums over the different local stabilizer configurations at site $i$. The jump operators $L^{(m)}_i$ are those defined Eqs.~\eqref{eq:ising_diss }-\eqref{eq:ising_diss2 } and Eqs.~\eqref{eq:4d_diss0}-\eqref{eq:4d_diss} (up to an orientation). If the local stabilizer configuration $m$ is not contained in the jumps in the main text, then $L^{(m)}_i = 0$. The rates $\gamma_m\geq 0$ and the local projectors on stabilizer configuration $P^{(m)}_i$ are chosen such that 
\begin{equation}
    \sum_m \left((L^{(m)}_i)^{\dag}L^{(m)}_i + \gamma_m (P^{(m)}_i)^{\dag}P^{(m)}_i\right) = \kappa.
\end{equation}
Therefore, applying the channel $\Lambda$ is equivalent to stochastically applying either a correcting step $\Lambda_r$ or a noise step $\Lambda_e$. The two steps are essentially update steps for the stabilizer configuration under the Glauber dynamics. Since the dynamics are only sensitive to the stabilizer configurations, we may use the states $\ketbra{\phi}{\phi}\in S$ to probe the spectrum of the Lindbladian in $\text{Span}(S)$, the numerical simulation becomes essentially classical.

The arguments above imply that 
we can directly include the ``do-nothing''  jumps into the definition of the Lindbladian $\hL$. This will not change the dynamics, i.e.~the subspace $\text{Span}(S)$ remains gapped and the Poissonian unraveling Eq.~\eqref{eq:poisson} becomes valid over the entire space $\mathcal{H}\otimes\mathcal{H}$.

\section{Fitting the autocorrelation with the sum of two exponential functions} \label{sec:2exp}
\begin{figure}
    \centering
    \includegraphics{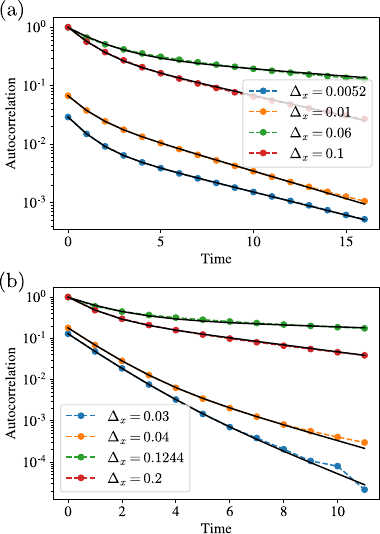}
    \caption{Autocorrelation functions for the 2D Ising model ($\kappa=1$), (a) the detailed balanced case and (b) the majority-vote case. The plot is in log scale on the y-axis. The black lines are obtained by fitting a sum of two exponential functions. Time is in units of $\delta t=(\kappa+\delta_x)^{-1}$.}
    \label{sm:fig:exp2}
\end{figure}
Here we show additional data supporting the fit of the 2D Ising autocorrelation function using a sum of two exponentials. The autocorrelation is plotted in Fig.~\ref{sm:fig:exp2} for some selected values of $\Delta_x$. It is clear from the autocorrelation that there are more than one time scales for the decay to happen. While a fit of a single exponential function can estimate the dominant decay time, a fit using a sum of two exponential functions gives a better resolution on the different decay time scales. 

\section{Equilibration time for  steady state sampling} \label{sec:equilibration}

\begin{figure}
    \centering
    \includegraphics[scale = 0.65]{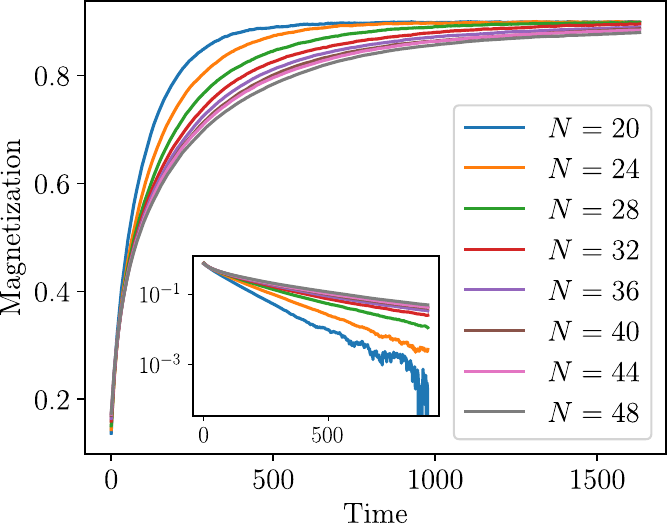}
    \caption{Relaxation to equilibrium magnetization. We show the magnetization dynamics $\Tr[\bar{Z}\rho(t)]$ (with $\bar{Z} = (\sum_i Z_i)/N^2$) for the 2D thermal Ising model ($\kappa = 1,\Delta_x = 0.02$) as it converges to one of the equilibrium value $\Tr[\bar{Z}\rho_{ss}]> 0$. The initial state is sampled from the ensemble where each spin has 60\% probability to be down and 40\% to be up. The result is averaged over $10^4$ trajectories. The inset shows the convergence to the stationary value with log scale on the y-axis.}
    \label{fig:expt}
\end{figure}

The mixing time for the thermalization (classical Glauber dynamics) of the 2D Ising model has been  well studied. In particular, at low-temperature, the ``true'' mixing time is known to scale exponentially with the system size due to spontaneous symmetry breaking~\cite{Thomas1989,Schonmann1987,randall:2006, temme:2013}. However, the equilibration time to sample from one of the symmetry-broken equilibrium states starting from any initial state is much less than that.

We can see this convergence explicitly using the channel evolution $\Lambda$ mentioned in the main text and Appendix~\ref{sec:channel} for an $N\times N$ 2D Ising model with detailed balance.  Starting from an ensemble that has an overall spin orientation that is far from equilibrium, the system converges to one of the equilibrium states rapidly, i.e.~the convergence is superpolynomial in $t$, and  the growth of the convergence time obeys slower than linear growth in system size $N$. In Fig.~\ref{fig:expt} we plot the convergence of the magnetization as a function of time for various system sizes. For Glauber dynamics in classical spin systems, the mixing time generally grows at least logrithmically with $N$~\cite{hayes:2007}. 

\begin{figure}
    \centering
    \includegraphics[scale = 0.65]{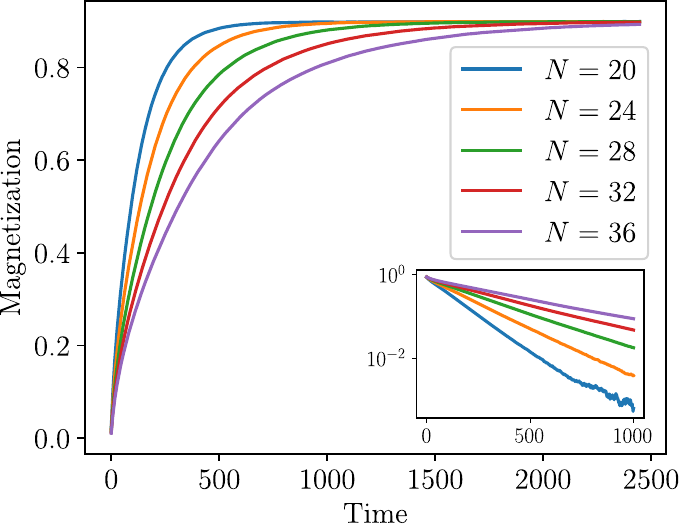}
    \caption{The relaxation starting from the infinite-temperature state. The setup is the same as in Fig.~\ref{fig:expt}. The inset shows the convergence to the stationary value with log scale on the y-axis.}
    \label{fig:expt5050}
\end{figure}

For the initial state with a completely random spin orientation (infinite temperature state), the convergence remains fast in time but the time it takes to relax appears to grow linearly or quadratically with $N$ (Fig.~\ref{fig:expt5050}). For a gapped primitive, reversible Lindbladian, the mixing time is $O(N^2)$~\cite{temme:2010,kastoryano:2013}, which is consistent with our numerics.
We expect the equilibration time  to be similar in the case of 4D toric code due to the analogous domain-wall-type dynamics.

\section{Connection to driven-dissipative phase transitions}\label{sec:driven-diss}

We have focused on thermal phase transitions in this work.    While thermal phase transitions are  caused due to a competition between energy and entropy, it is known that dissipative systems can undergo \textit{non-equilibrium} phase transitions which arise due to a different mechanism: The competition between a quantum coherent drive and dissipation. These are called \textit{driven-dissipative} phase transitions \cite{diehl2008,  maghrebi2016,young2020, keeling2013, rossini2018,  ciuti2018, lieu2020, eisert2017, cirac2012}. The dynamics of such systems is more ``quantum'' in the sense that we need to simulate the full quantum Hilbert space within the trajectory approach (unlike the thermal transitions above which are efficiently simulable on a classical computer). To our best knowledge, all examples of driven-dissipative phase transitions  arise due to spontaneous symmetry breaking. Is it possible to achieve a driven-dissipative topological phase transition? And can this be done in less than 4D? Here we briefly review the driven-dissipative phase transition in the transverse-field Ising model and speculate on a topological model which might exhibit a transition, albeit in 4D. 

Consider the transverse-field Ising Hamiltonian in the presence of dissipation:
\begin{equation}
H = - J \sum_{\langle i j \rangle } X_i X_j - h \sum_i Z_i, \qquad L_i = \sqrt{\gamma} \sigma_i^-,
\end{equation} 
where $\sigma_i^-$ is the lowering operator in the $Z$ basis \cite{overbeck2017,keeling2013}. This can be viewed as the rotating-frame Hamiltonian of a lattice of spins in the presence of a coherent drive \cite{keeling2013}. It is believed that this model has a phase transition in 2D and higher: For $J/h \ll1$ the model is in a trivial paramagnetic phase; for $J/h \sim 1$ and $\gamma/h \sim 1$, the drive causes the steady state to spontaneously break the symmetry \cite{overbeck2017}. This transition is most easily understood within the quantum jump picture:  The jump operators want to evolve the system to a state with all spins pointing down, but the non-Hermitian effective Hamiltonian arising from the nearest-neighbor coupling ($J$) will cause the spins to rotate. The competition between these two processes will lead to a phase with net magnetization in $X$ when the drive crosses a certain critical strength.

Working by analogy, we speculate that the following 4D model might exhibit a driven-dissipative \textit{topological} transition:
\begin{equation}
H =  -J \sum_{\vec{v}, \hat{e}}   S_{\vec{v},\hat{e}} - J \sum_{\vec{v}, \hat{c}}   S_{\vec{v},\hat{c}} - h \sum_i Z_i, \qquad L_i = \sqrt{\gamma} \sigma_i^-
\end{equation} 
where  the stabilizers $S$ are defined in Sec.~\ref{sec:4d_tc}. (The terms with a prefactor $J$ are just the 4D toric code Hamiltonian.) Again we expect a trivial paramagnetic phase  for $J/h \ll1$  since the dissipation acts as a zero-temperature bath in this limit. Nevertheless, for larger values of $J$ the Hamiltonian evolution could start to cause the (generally mixed) steady state to  acquire a non-zero topological order parameter. An interest direction for future work involves characterizing the phases of such a model.

\bibliography{toric_code.bib}
\bibliographystyle{apsrev4-1}

\end{document}